\documentstyle[sprocl,epsfig]{article}
\def\be{\begin{equation}}
\def\ee{\end{equation}}
\def\ba{\begin{eqnarray}}
\def\ea{\end{eqnarray}}

\begin{document}
\title{%
PREDICTIONS FOR FERMION-PAIR PRODUCTION AT LEP
}
\author{%
P. CH. CHRISTOVA\footnote{Supported by Bulgarian foundation for
Scientific Research with grant $\Phi$--620/1996.}  
}
\address{%
Faculty of Physics, Bishop Konstantin Preslavsky Univ., 
\\ Shoumen, Bulgaria
\\ E-mail: penka@main.uni-shoumen.acad.bg
} 
\author{M. JACK, S. RIEMANN, T. RIEMANN
}
\address{DESY Zeuthen, Platanenallee 6,
\\ D-15738 Zeuthen, Germany
\\ E-mails: jack@ifh.de, riemanns@ifh.de, riemann@ifh.de}
\maketitle\abstracts{ 
The status of predictions for 
fermion-pair 
production at LEP is summarized
with emphasis on LEP2 energies and on the physics interest there.
Some numerical comparisons with other programs 
are performed.  
We also present first results of a semi-analytical recalculation of photonic
corrections with acollinearity cut in the {\tt ZFITTER} approach.
}
\section{Introduction
}
Since RADCOR'96, the era of high precision measurements of 
fermion-pair 
production at the $Z$ resonance has been finished.
For latest results on precision tests of the Standard Model physics,
see 
\cite{Barc-Hollik:1998x,Barc-Teubert:1998x,Barc-Gambino:1998aax}.    
At LEP2 the $Z$ resonance region is left behind with counting rates
being a factor of thousand or so smaller.
Correspondingly, precision demands for the predictions are weakened;
see Table \ref{accuracies}. 
A summary of Standard Model predictions for fermion-pair 
production at LEP2 may be found in \cite{Boudjema:1996qgx}.  
The hard photonic bremsstrahlung is no longer suppressed, 
the initial-final state interference corrections reach the order of a percent,
and weak virtual corrections are also enlarged and show subtle 
dependences on kinematics.
The interest in fermion-pair production at LEP2 focuses on searches
for virtual signals from New Physics.
First results have been reported recently.

 \begin{table}
\begin{center}
\vspace*{-0.5cm}
\caption{
Some observables for fermion-pair 
production at LEP. Experimental accuracies are given in square brackets.
For LEP1 the total experimental error is shown, for LEP2 the estimated 
statistical error per experiment based on present efficiencies. 
\label{accuracies}
}
\vspace{0.3cm}
 \renewcommand{\arraystretch}{1.2}
\begin{tabular}[tbh]{|c|cc|cc|cc|}
\hline
   &  \multicolumn{2}{|c}{$\sqrt{s} = M_Z$}
   &  \multicolumn{4}{|c|}{$\sqrt{s} = 189$ GeV}
\\ 
   &  \multicolumn{2}{|c}{unfolded, $Z$ only}
   &  \multicolumn{2}{|c}{no cut}
   &  \multicolumn{2}{|c|}{$\sqrt{s'/s} > 0.85$}            \\
\hline
$\sigma_{had}  $ [pb] & 41~486 & [0.14\%]
                      & 102.1  & [0.8\%]  & 22.1 & [1.7\%] \\ \hline
$\sigma_{\mu}  $ [pb] & 1~996  & [0.17\%]
                      & 9.43   & [3.0\%]  & 3.07 & [4.6\%] \\ \hline
$R_{b}=\sigma_b/\sigma_{had}$
                      & 0.2156 & [0.34\%]
                      & 0.184  & [2.0\%]  & 0.1651 & [3.6\%] \\ \hline
$A_{FB}^{\mu}$        & 0.0164 & [0.0013]
                      & 0.228  & [3.1\%]  & 0.585 & [4.0\%] \\ \hline
$N_{had}$             &  {14.800~k}  &
                      & 16700  &     & 3600 & \\
$N_{lep}$             & \multicolumn{2}{|l}{1.600~k ($e + \mu + \tau$)}
                      & \multicolumn{2}{|l}{1150 ($\mu$)}
                      & \multicolumn{2}{|l|}{~~450 ($\mu$)}
\\ \hline
\end{tabular}
\end{center}\end{table}

\subsection{The $\gamma Z$ Interference and $M_Z$}
In a model-independent approach, there is a strong correlation between
$M_Z$ and the $\gamma Z$ interference term $\cal J$ in total cross-sections:
\cite{Leike:1991pq}
\ba
\sigma_T^0(s) \sim \frac{\alpha_{em}^2(M_Z)}{s} + 
\frac{{\cal R}s+{\cal J}(s-M_Z^2)}{|s-M_Z^2+iM_Z\Gamma_Z(s)|^2} .
\label{smeq}
\ea
This correlation has been studied at LEP energies.
The hadron production data allow to deduce from (\ref{smeq})
(~\cite{Wynhoff:1998x} and references therein):
\ba
M_Z &=& 91\,188 \pm 3 \pm 2.7~~ \mathrm{MeV}.
\label{smmz}
\ea
When determined from the $Z$ peak data alone, the error in
(\ref{smmz}) is $\pm 3 \pm 13$ MeV. 
The Standard Model fit yields $M_Z = 91186.7 \pm 2.1$ MeV 
\cite{Vanc-Gruenewald:1998x}
where the $Zf{\bar f}$ couplings and thus $\cal J$ in (\ref{smeq}) are
fixed.
The very good agreement of the two fit procedures is a valuable test
of the Standard Model.\cite{Riemann:1997tjx}  
\subsection{Virtual Effects due to New Physics
\label{other}
}
LEP2 has some potential for the observation of new
virtual effects in the $2f$ final state. 
Recent results are summarized in 
\cite{Wynhoff:1998x,Pieri:980723x} and references therein:
\begin{itemize}
\item
Heavy neutral $Z'$ bosons may be searched for in two
respects.\cite{Leike:1992ufx,Chiappetta:1996avx}
At the $Z$ peak, 
limits on a $ZZ'$ mixing angle may be derived, typically
$\theta_{M} < $ O(few parts per mil). 
While, at LEP2 limits on the $Z'$ mass are obtained in the range
$M_{Z'}> 270 -- 820$ GeV depending on the models studied.
\item
A limit on the energy scale $\Lambda$ at which contact interactions
could appear is $\Lambda > 4 - 10$ TeV.
Typical limits from atomic parity violation searches are $\Lambda
> 15$ TeV.
They are not sensitive to
the $\cal P$ conserving $VV, AA, LL+RR, LR+RL$ type models.
\item
Leptoquarks, and also  sneutrinos and squarks from
supersymmetric theories with $\cal R$-parity breaking 
may be exchanged in addition to $\gamma$
and $Z$.
E.g. the leptoquark mass limits, $m_{LQ} > 120 - 430$ GeV,
are for some models competetive with direct searches. 
\end{itemize}

\section{Realistic Observables
\label{sec-real}
}
Cross-sections, $\sigma^0(s)$, and asymmetries, $A^0(s)$, are called 
{\em improved Born approximations} or {\em unfolded observables} if they are
free of photonic corrections.
{\em Realistic observables}, $\sigma(s)$, contain the photonic corrections.
Examples of numerical programs for the
calculation of realistic observables are
{\tt ALIBABA},\cite{Beenakker:1991mb}
{\tt BHM},\cite{Burgers:BHM}
{\tt KORALZ, KK},\cite{Jadach:1994yvx,Jadach:1998-235x,Jadach:1998-253x}
{\tt TOPAZ0 v.4.3},\cite{Montagna:1998kpx}
{\tt ZFITTER v.5.14}.\cite{zfitter:v5.14x}
In the {\tt ZFITTER} approach,
\cite{Bardin:1992jcx,Bardin:1991dex,Bardin:1991fux,Bardin:1989dix}
we calculate:
\ba
\sigma(s) \sim \int \frac{ds'}{s}~ {\sigma^0(s')}~ \rho(s'/s) .
\label{siggen}
\ea
Here, $s'=m^2_{f\bar f}$ is the invariant mass of the 
fermion pair 
and $\rho$ some radiator. 
With {\tt ZFITTER}, a {\em one-dimensional} numerical integration is
performed for three different kinematical treatments of photonic corrections: 
(i) no cut,\cite{Bardin:1989cwx}
(ii) cuts on $s'$ and on the scattering angle $\vartheta$ of one
fermion,\cite{Bardin:1991fux,Bardin:1991dex} 
or  
(iii) cuts on the fermions' acollinearity angle, on
  their energies $E^{f}=E^{\bar f}$, and on $\cos \vartheta$; see Section
       \ref{sec-acol}.  
The effective Born cross-section, {$\sigma^0(s')$}, may also be chosen
according to following approaches:
(A) Standard Model,\cite{Bardin:1989dix}
(B) Model Independent,\cite{Bardin:1992jcx}
(C) Others (see Section \ref{other}).
Since the last comprehensive review \cite{Bardin:1995aax}
many careful comparisons have been undertaken.
For LEP1 applications the theoretical accuracy is now
considered to be excellent and sufficient for data samples of $O(10^7)$;
for fermion-pair 
production see \cite{Bardin:1998nmx,DBGPLEP98}
and  for (wide-angle) Bhabha scattering at LEP1 see \cite{Beenakker:1997fi}.
A recent overview
on precision physics at LEP is \cite{Montagna:1998spx}.
For the Monte-Carlo
approach see \cite{Jadach:1998aax,Jadach:1998-235x,Jadach:1998-253x}.
We would also like to recommend to
regularly consult home pages or {\tt afs} accounts 
of the Dubna/Zeuthen group (e.g. \cite{home-Riemann} or \cite{afs-Bardin}), 
the Krakow/Knoxville group (e.g. \cite{home-Jadach}), 
the Torino/Pavia (e.g. \cite{home-Passarino}) 
group, and also of the LEPEWWG (~\cite{home-LEPEWWGx}).
\section{The $Zb\bar b$ Vertex at LEP2
\label{zbb}
}
\begin{figure}[tb] 
\begin{center} 
  \mbox{%
  \epsfig{file=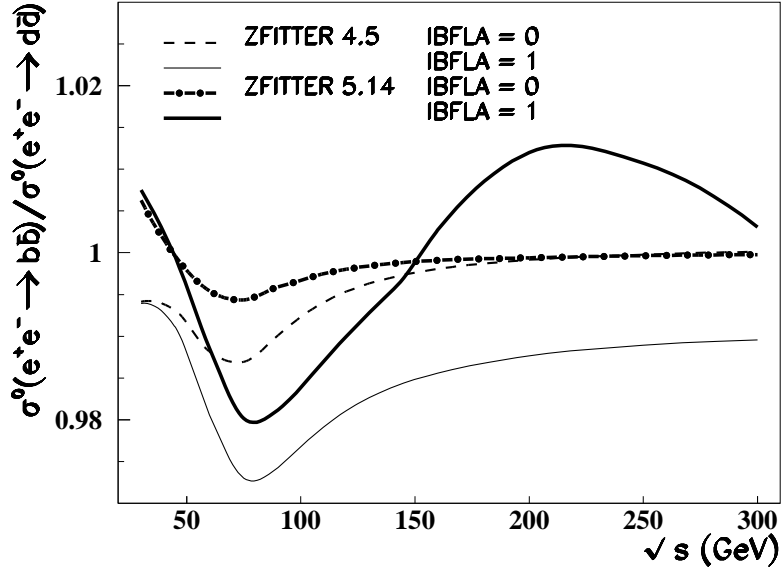
          ,height=8.cm  
         }%
  }
\caption 
{
Ratios of  improved Born cross-sections for $b\bar b$ and $d \bar d$
production from {\tt ZFITTER} v.4.5 (1992) and v.5.14 
(1998); the latter has the correct $t$ mass dependence at LEP2.    
\label{fig-zbb} 
}
\end{center}
\end{figure}

An instructive example for different behaviour of the weak corrections on and
off the $Z$ peak is $b\bar b$ production.
The corrections differ from those to $d\bar d$ production due to the huge
$t$-quark mass and may be described at LEP1 with formulae derived for the $Z$
width.\cite{Akhundov:1986fcx,Bardin:1992jcx,Bardin:1995aax}
At higher energies, there are further contributions of similar size from the
$\gamma b \bar b$ vertex and from the $W^+W^-$ box.
Further, one has to take into account the $s$-dependence of the vertices and
for the box also the angular dependence.
The net effect is taken into account in {\tt ZFITTER} since v.5.12
and is shown in Fig. \ref{fig-zbb}. 
It may be switched off with flag {\tt IBFLA}=0.
It amounts to about 2--4 \%
and is thus of the order of the statistical error; see Table \ref{accuracies}.

\section{Photonic Corrections with Acollinearity Cut: Comparisons
\label{comparisons}
}
 
\begin{figure}[tb] 
\begin{center} 
\mbox{ 
       \epsfig{file=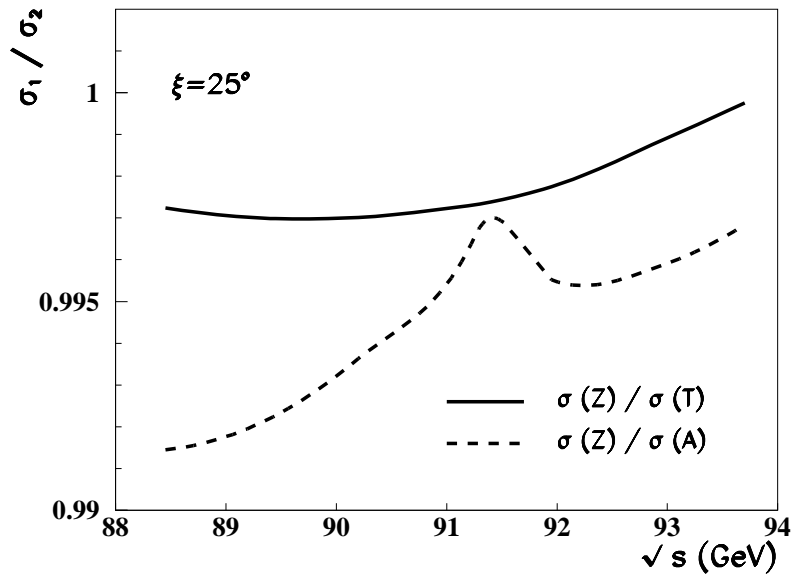,
                    height=8.cm,%
                   clip=%
}} 
\caption []
{
Ratios of $s$-channel contributions to Bhabha scattering at LEP1: 
{\tt ZFITTER} v.5.14 versus {\tt TOPAZ0} and {\tt ALIBABA}.
\label{sigma9} 
} 
\end{center} 
\end{figure} 

\begin{figure}[tb] 
\begin{center} 
\vspace*{-.2cm}
  \mbox{%
  \epsfig{file=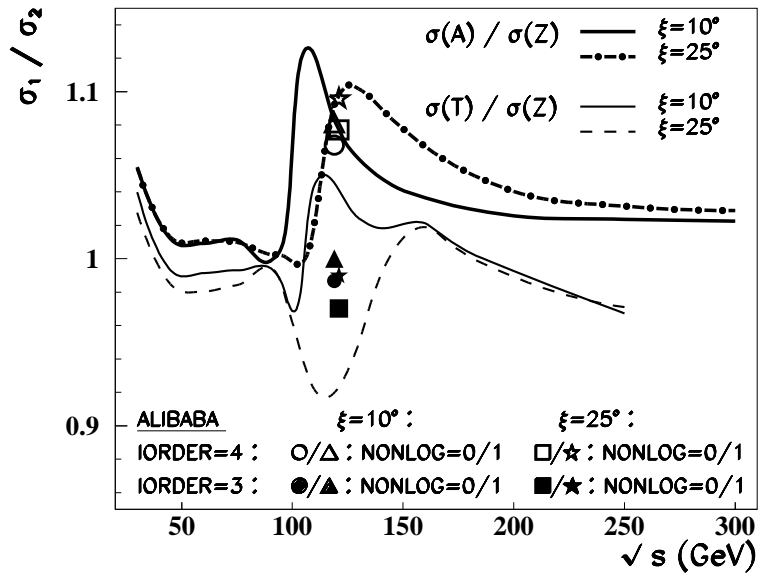
          ,height=8.cm  
         }%
  }
\caption []
{
Numerical comparison of {\tt ALIBABA} v.2, {\tt TOPAZ0} v.4.3, and
{\tt ZFITTER} v.5.14 as functions of $s$ with two settings of
acollinearity cut $\xi$ ($40^{\circ} < \theta^{\bar f} < 140^{\circ}$). 
At $\sqrt{s}$ = 120 GeV, we also show data points for variations of
flag settings (i,n,m) in {\tt ALIBABA} as discussed in the text.
\label{comp4a} 
}
\end{center}
\end{figure}

For the analysis of experimental data the treatment of kinematical
cuts on the final particles' phase space is of utmost importance.
A variety of numerical comparisons for LEP1 may be found in
\cite{Bardin:1992jcx}. 
Later, the $s'$-cut was studied for LEP1 in \cite{Bardin:1995aax}.
Numerical results with acollinearity cut are given in
Table 3 of \cite{Beenakker:1997fi} for the $s$-channel part of Bhabha
scattering.
In Fig. \ref{sigma9}, we compare this  with {\tt ZFITTER}
v.5.14. and get very good agreement.
For LEP2, the  $s'$-cut is estimated to be `under control' in
\cite{Boudjema:1996qgx} while a warning was given there
that `the agreement between {\tt TOPAZ0} and {\tt ZFITTER} somehow
degrades when implementing an acollinearity cut'.
However, one should
mention here that both programs were originally designed for applications
around the $Z$ resonance and using them at higher energies deserves
dedicated checks and, if necessary, further improvements.  
For a wider energy range, including LEP2, a comparison of
$\sigma_T$ and $A_{FB}$ from {\tt ALIBABA} v.1 (1990) and {\tt
ZFITTER} v.4.5 (1992) shows deviations up to 10\% for the
acollinearity cut option at energies above
the $Z$ resonance.\cite{Riemann:199200}
Within the present study, we add some numerical comparisons of
{\tt ALI\-BA\-BA, TOPAZ0}, and {\tt ZFITTER} over a wide energy
range.\cite{DESY98-184x} 
{\tt ALIBABA} v.2 (1990) was used with the default settings and in
{\tt ZFITTER} v.5.14 we modified one flag ({\tt PHOT2}=2). 
{\tt TOPAZ0} v.4.3 was run in accordance with {\tt ZFITTER} v.5.14.
Fig. \ref{comp4a} shows cross-section ratios as functions of $s$
with parameter  $\xi$, the cut on the maximal acollinearity angle
between the fermions.
The gross features of the 1992 comparison are retained  with the new
program versions. 
Between about 100 GeV and 200 GeV, 
the deviations in the predictions from different programs are huge and
heavily depending on  $\xi$, here shown for  $\xi=10^{\circ},25^{\circ}$.
The ratios stabilize at higher (or smaller) energies.
In addition, we show at 120 GeV selected predictions arizing from a
variation of flags ({\tt{IORDER,NONLOG,IFINAL}}) in {\tt ALIBABA}:
upper ones at (4,n,m), lower ones at (3,n,m), with n=0,1, m=1,2 (best
choice: (4,1,2)).   
This visualizes the strong dependence of predictions on the 
details of the theoretical input chosen, e.g. 
the treatment of higher order contributions or
the correct inclusion of non-logarithmic $O(\alpha)$ corrections.
Evidently, the largest deviations arise from the radiative return of 
$\sqrt{s'}$ to the $Z$ resonance due to hard initial state
radiation. 
Interest in the high energy part of the data anyhow means to cut 
this away and so there shouldn't be a serious problem.
If instead one is interested in the radiative return, one has to be
 concerned about accuracies.
These observations confirm similar statements from other
studies.\cite{Montagna:1997jtx,Placzek:1998x}

\section{Semi-analytical Approach to the Acollinearity Cut
\label{sec-acol} 
}
\begin{figure}[tb] 
\begin{center} 
\vspace*{-1.2cm}
  \mbox{%
  \epsfig{file=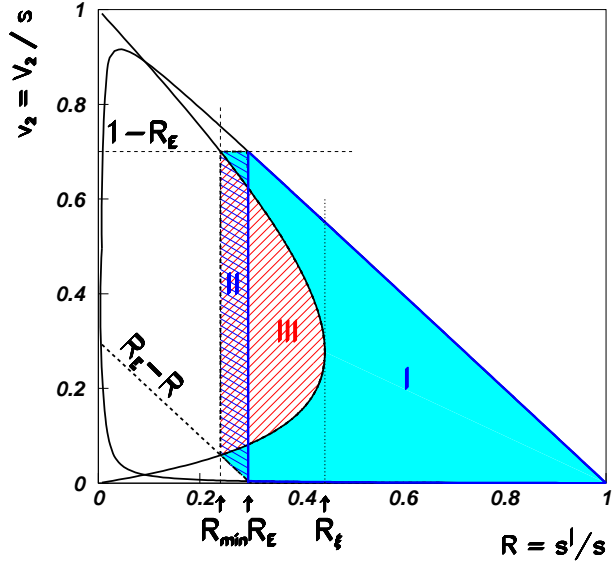
          ,height=9cm  
          ,width=9cm   
         }%
  }
\caption[]{\label{dalitz} 
 Phase space with acollinearity cut 
and (lower) muon energy cuts. 
}
\end{center}
\end{figure}

A sketch of the analytical formulae with acollinearity cut coded in 
{\tt ZFITTER} is given in \cite{Bilenkii:1989zgx}. 
Since some simplifications were made which were intended for applications near
the $Z$, we now recalculate the corresponding $O(\alpha)$ corrections. 
The kinematics were derived in \cite{Passarino:1982zp}. 
One has to perform a three-fold analytical integration of the squared
matrix element over one photonic angle, over the invariant mass
$x$ of (fermion+photon) in the cms, and over the fermion's scattering angle. 
The Dalitz plot is shown in Fig.~\ref{dalitz} ($v_2=x$).

The cross-section is the sum over three regions in phase space:
\ba
\sigma(s) = \left[ \int_{\mathrm{I}} + \int_{\mathrm{II}}  
- \int_{\mathrm{III}} \right] ~ ds' ~ dx ~ 
d\cos\vartheta \frac{d\sigma(A)}{ds' dx d\cos\vartheta} .
\label{sig}
\ea
Parameter $A=A(s'/s)$ has different meaning in these regions: 
\ba
A_{\mathrm{I}} &=& 1,
\\
A_{\mathrm{II}} &=& (1+R-2R_E)/(1-R), 
\\
A_{\mathrm{III}} &=& [1 - R(1-R_{\xi})^2/(R_{\xi}(1-R)^2)]^{1/2}, 
\label{A}
\ea
with $R_E = 2E_{min}/\sqrt{s}$, 
$R_{\xi} = (1-\sin(\xi/2))/(1+\sin(\xi/2))$, and $R=s'/s$. 
We have (preliminary) results for the numerically largest initial state 
corrections and see deviations from the old results in the hard
bremsstrahlung $\sigma^{hard}$.  
For the total cross-section, e.g., the analytical formula with cuts on
acollinearity and minimal fermion energy is remarkably compact
for the full angular acceptance ($c=1$). 
In each of the three regions, it is:
\ba
\sigma_T^{hard}(s,\xi, E_{min}) &=& 
\frac{3\alpha}{4\pi} Q_e^2 ~
\int dR ~ \sigma_T^0(s')
~\rho_T(R,A), 
\ea
\ba
\rho_T (R,A)  &=& 
\left(A+\frac{A^3}{3}\right) \frac{1+R^2}{1-R}
\left( \ln\frac{s}{m_e^2}-1\right) + (A-A^3) \frac{{\cal B}R}{1-R},
\label{sigtA}
\ea
with ${\cal B}=2$.
For $A \to 1$, the phase space regions II and III do not contribute and 
(\ref{sigtA})
approaches the well-known result derived in \cite{Bonneau:1971x}.
The additional contributions from final state radiation and the initial-final
state interference to $\sigma_T$ (and also those to $\sigma_{FB}$) may be
found in \cite{Bardin:1989cwx} for $A=1$. 
The other generalizations for  $A \neq 1$ will be published elsewhere. 
Some analytical formulae for the final state corrections are given in
\cite{Montagna:1993mfx}.
We have to mention that, differing from (\ref{sigtA}), 
the coding in {\tt ZFITTER} corresponds to  ${\cal B}= 4/3$
if one looks there into the limit $c=1$.
The resulting numerical deviations are 
typically of the order of 0.5\% to 2\%. 
They do not lead to drastical
improvements of the comparisons shown in Section \ref{comparisons}.
\section{Summary}
We gave a brief overview on recent developments of predictions for 
fermion-pair 
production at LEP and on some physics results from LEP2.
Predictions at LEP2 energies when applying an
acollinearity cut are discussed in more detail.
First results of a recalculation of the analytical formulae for
hard photon corrections in {\tt ZFITTER} with acollinearity cut are
presented. 
The numerical effects are of the order of 1\%.  
The need of further investigations is stressed.

An extended version of this contribution is \cite{DESY98-184x}.
\section*{Acknowledgments}
We thank G.~Passarino for assistance by delivering numbers from 
{\tt TOPAZ0} for Fig. \ref{comp4a},
W.~Beenakker for making available the actual version of  
{\tt ALIBABA}, and S.~Jadach for discussions and informing us on comparisons
of {\tt KK} with {\tt ZFITTER}.
T.R.  would also like to thank J.~Sol\`a and the organizing committee at 
Universitat Aut\`onoma de Barcelona for the enjoyable running of RADCOR'98.

\end{document}